\documentclass{interact}

\usepackage[T1]{fontenc}
\usepackage[utf8]{inputenc}
\usepackage{mathpazo}
\usepackage[english]{babel}
\usepackage{csquotes}
\usepackage{microtype}
\usepackage{booktabs}
\usepackage{amsmath,amssymb}
\usepackage{musicography}
\usepackage{graphicx}
\usepackage{xcolor}
\usepackage{listings}
\usepackage{hyperref}
\hypersetup{pdfborder = {0 0 0}}
\usepackage[backend=biber, style=authoryear, useprefix=false]{biblatex}
\usepackage{endnotes}
\let\footnote=\endnote
\makeatletter
\AtBeginDocument{\toggletrue{blx@useprefix}}
\AtBeginBibliography{\togglefalse{blx@useprefix}}
\makeatother
\newcommand{\spn}[2]{\text{#1}_{#2}}

\newcommand\noteE{\spn E3}
\newcommand\noteF{\text{F}^\#_3}
\newcommand\noteG{\spn G3}
\newcommand\noteA{\spn A3}
\newcommand\noteB{\spn B3}
\newcommand\notec{\spn C4}
\newcommand\noted{\spn D4}
\newcommand\notee{\spn E4}
\newcommand\notef{\text{F}^\#_4}
\newcommand\noteg{\spn G4}

\newcommand{\MM}{\textsc{m}}
\newcommand{\TT}{\textsc{t}}

\newcommand*{\concat}{\mathbin{\raisebox{-.2em}{$\smallsmile$}}}
\begin{document}

\title{
    Algo Pärt: An Algorithmic Reconstruction of Arvo Pärt's \emph{Summa}}

\author{\name{
    Bas Cornelissen
    }
    \affil{
        Institute for Logic, Language and Computation,\\University of Amsterdam, The Netherlands
    }
}

\received{Completed: February 9, 2023\\Preprint v1: March 27, 2026}

\maketitle

\begin{abstract}
    Arvo Pärt is one of the most popular contemporary composers, known for his highly original \emph{tintinnabuli} style. 
    Works in this style are typically composed according to precise procedures and have even been described as algorithmic compositions. 
    To understand how algorithmic Pärt’s music exactly is, this paper presents an \emph{analysis by synthesis}: it proposes an algorithm that almost completely reconstructs the score of \emph{Summa}, his “most strictly constructed and most encrypted work,” according to Pärt himself in 1994. 
    The piece is analyzed and then formalized using so-called tintinnabuli processes.
    An implementation of the resulting algorithm generates a musical score matching Summa in over 93\% of the notes. 
    Due to interdependencies between the voices, only half of the mistakes (3,5\%) need to be corrected to reproduce the original score faithfully. 
    This study shows that \emph{Summa} is a largely algorithmic composition and offers new perspectives on the music of Arvo Pärt.
\end{abstract}


\section{Introduction}

Music and algorithms share a long history, but rarely has their marriage been as fruitful as it has been in the hands of the Estonian composer Arvo Pärt.\nocite{Edwards2011}
According to one study, Pärt was the most frequently performed contemporary composer from 2011 until 2019.%
\footnote{
    This is based on data released by Bachtrack, a classical music website that tracks many thousands of concerts every year. 
    The website annually releases statistics about concert performances, including the most performed classical composer.
    In the year 2018 (\url{bachtrack.com/classical-music-statistics-2018}), these statistics were based on almost 20.000 concerts, in which Pärt was the top contemporary composers, as he had been since 2011 (see \url{bachtrack.com/classical-music-statistics-2017}). 
    In 2019, John Williams came out first, with Pärt second.
}
Not only is his music popular, but it is also highly original.
In the 1970s, Pärt developed a unique compositional technique, known as \emph{tintinnabuli}, that is deeply algorithmical due to its use of numerical procedures.
The main melody may, for example, walk down a scale, moving one step further with every measure.
Alternatively, it may be determined by the text: in his \emph{Missa Sillabica}, the number of syllables in a word determines the melody for that word.
Examples such as these raise the question \emph{how} algorithmic Pärt's music precisely is.
Can all notes in a score be explained by formal procedures? 
And when does the composer deviate from those, if at all?

To address such questions, I propose a type of computational music analysis \parencite[cf.~][]{Anagnostopoulou2010} that one could call \emph{analysis by synthesis}.
Motivated by the idea that one cannot understand what one cannot create, the aim is to implement an algorithm that reconstructs as much of a score as possible.
By measuring the \emph{reconstruction error}, the number of errors in the reconstructed score, one can evaluate the algorithm.
In practice, such an analysis is an iterative process in which one successively refines the rules to further reduce the reconstruction error.
As the error decreases, the explanatory power of the algorithm increases, until adding new rules no longer seems to be theoretically productive.
Adding a rule that explains only a single note, for example, is not very productive and similar to `overfitting' a mathematical model.
But up to that point, the algorithm provides an answer to a central question of musical analysis: how does the piece work?

The idea of using algorithms to analyze Pärt's music is not new.%
\footnote{
    I found two conference papers that use small fragments of Pärt's compositions as examples in a live coding setting \parencite{Bertram2014, Ruthmann2010}.
    \textcite{Kramer2015} cites a script by Christopher Ariza and Michael Scott Cuthbert that generates a score for Pärt's \emph{Pari Intervallo}, which can indeed be found in an old release of music21: \url{github.com/changtailiang/music21/blob/master/music21/composition/phasing.py}.
    David Cope appears to have discussed \emph{Cantus in Memoriam Benjamin Britten} in a course on computer-assisted composition in 2008.
    \textcite{DePaivaSantana2012} presented a poster that modelled \emph{Spiegel im Spiegel} in OpenMusic \parencite[see also][]{Shvets2014a}.
    Outside the academic literature, Guy Birkin in 2015 released the album \emph{Tintinnabuli Mathematica vol. I} with music generated in Mathematica using tintinnabuli rules and number sequences. 
    He explains the process in a blog post available at \url{aestheticcomplexity.wordpress.com/2011/11/11/programming-arvo-part}.
}
\textcite{Shvets2014} describes multiple constructions commonly found in the work of Pärt using concepts borrowed from programming languages, such as loops.
\textcite{Shvets2014a} then went on to implement several models of Pärt's compositions.
In a more formal analysis, \textcite{Roeder2011} proposes to understand Pärt's compositional procedures as musical transformations \parencite[cf.][]{Lewin1987}.
His analysis effectively results in several (functional) programs that model certain aspects of Pärt's music.
This paper takes these ideas one step further by first formalizing a piece, then implementing an algorithm to reconstruct the full score, and finally quantitatively evaluating that against the original: a complete analysis by synthesis.

Our case study looks at \emph{Summa}.
This piece was written in 1977, one year after Pärt wrote his first piece in tintinnabuli style (\emph{Für Alina}).
\emph{Summa} is best known as a composition for mixed choir or solo voices but was originally written for two voices (tenor and bass) and six instruments \parencite{Hillier1997}.
It has since been adapted for many instrumental combinations, from string quartet to trombone quartet.
The composition is intricately structured, but many of the underlying regularities  will escape notice when listening to a performance, or even when studying the score.
Indeed, some of the procedures identified in this paper seem to have escaped previous analyses of \emph{Summa} \parencite{delaMotte-Haber1996, Hillier1997, Patrick2011}.
Arvo Pärt may have anticipated this when he wrote:
\begin{quote}
    I have developed a highly formal compositional system in which I have been writing my music for 20 years. 
    In this series, \emph{Summa} is the most strictly constructed and most encrypted work.
    The encryptions are found in many layers of the score.%
    \footnote{
        \label{fn:parts-comment}
        My translation. It is instructive to read his comment in full:
        \begin{quote}
            \footnotesize
            Ich habe große Schwierigkeiten, wenn ich meine Werke kommentieren soll. 

            \bigbreak\noindent
            Ich bin dafür, daß zwischen Wort und Musik ein besonders behutsames Verhältnis sein muß. Wir müssen der Musik eine Chance geben, sich allein auszudrücken. Wörter treiben die Musik in die Enge. Und auch die Musik neigt dazu, sich von Wörtern abhängig zu machen. Ich sehe in dieser »überkommunizierten« Gesellschaft Gefahr für die Existenz der Musik.
        
            \bigbreak\noindent
            Ich muß in mir Raum frei lassen für Musik, und wenn dieser Raum mit Worten besetzt wird, bleibt mir kein Bedürfnis, mich mit Musik auszudrücken — und umgekehrt: wenn ich ein Musikwerk geschrieben habe, bleibt nichts mehr mit Worten zu sagen übrig.
            \begin{align*}
                \star \quad \star \quad \star
            \end{align*}
        
            \noindent
            Ich habe ein hochformalisiertes Kompositionssystem entwickelt, in dem ich seit 20 Jahren meine Musik schreibe. In dieser Reihe ist \emph{Summa} das strengstgebaute und verschlüsselste Werk. Die Verschlüsselungen finden sich in vielen Schichten der Partitur.
        
            \bigbreak
            \hfill(Berlin, den 15.6.1994)
        \end{quote}
    }
    \parencite{Part1996}
\end{quote}
I read this as an invitation to decrypt \emph{Summa}.
But first, let me introduce the tintinnabuli style and the terminology that \textcite{Hillier1989, Hillier1997} developed to describe it---some of which Pärt himself has adopted.

\section{Tintinnabuli}

At the heart of the tintinnabuli style lie two voices: a melodic voice or \emph{\MM-voice} and an accompanying tintinnabuli voice or \emph{\TT-voice}.
The \MM-voice is usually diatonic and tends to move in steps around a pitch center.
It is sometimes freely composed, but more often constructed according to numerical procedures.
The accompanying \TT-voice is even more constrained.
It can only use notes from a central \emph{tintinnabuli triad}, and is determined by its \MM-voice according to a strict procedure, such as always using the first note in the triad above the \MM-voice.
The resulting texture is often homophonic, further emphasizing the unity of the two voices.
For Pärt, the \MM- and \TT-voice are indeed much more than a compositional technique.
The \MM-voice ``signifies the subjective world, the daily egoistic life of sin and suffering; the \TT-voice, meanwhile, is the objective realm of forgiveness'' \parencite[p.~96]{Hillier1997}.
Their duality is only appearance: they really are a ``twofold single entity,'' as has been summarized in the equation $1 + 1 = 1$.

\begin{figure}[t]
    \centering
    \includegraphics[width=\textwidth]{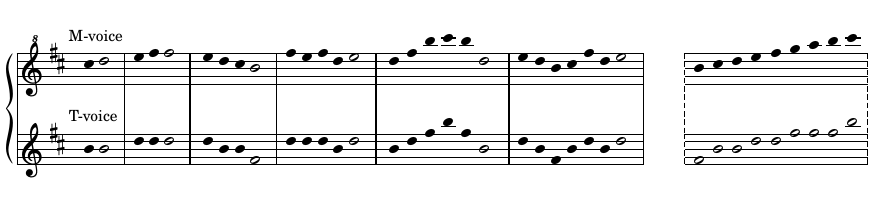}
    \caption{
        \textbf{Excerpt of \emph{Für Alina}} (mm.~2--7).
        This piano piece was Arvo Pärt's first work in his tintinnabuli style. 
        The right hand plays a melodic voice (\MM-voice) that mostly moves stepwise, and which is freely composed.
        The left hand has the tintinnabuli voice (\TT-voice), that is restricted to notes from the B-minor triad.
        The relation between the two voices is shown on the right: the \TT-voice plays the highest triad note below the \MM-voice, but one octave lower.
    }
    \label{fig:fur-alina}
\end{figure}
\begin{figure}[t]
    \centering
    \includegraphics[width=\textwidth]{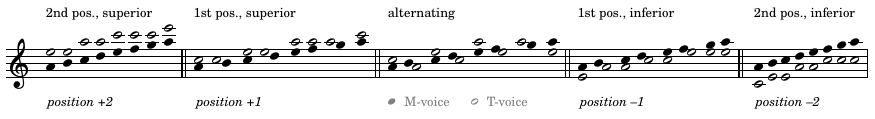}
    \caption{
        \textbf{Tintinnabuli positions} for an A minor triad. 
        Solid notes show the A minor scale as a melodic voice, and open notes show tintinnabuli voices in the five different positions introduced by \textcite{Hillier1997}.
        His terminology is shown above the staff, the numbering used in \autoref{sec:synthesis} below it.
    }
    \label{fig:positions}
\end{figure}

To clarify the terminology, let's consider two famous examples (see e.g.~\cite{Hillier1997} for more extensive analyses).
\autoref{fig:fur-alina} shows an excerpt of the piano piece \emph{Für Alina} (1976).
The piece is composed around the B minor tintinabulli triad, and the tonal center of B is reinforced by a low pedal note not shown in this excerpt.
The right hand plays the \MM-voice, and the left hand plays the \TT-voice in the same rhythm, using only notes from the tintinnabuli triad.
The relation between the two voices is simple: the left hand plays the highest triad note below the melody, but one octave lower.
Pärt deviates from this only once, when the \TT-voice plays a C\musSharp{} (in bar 11; not shown). 
This special event is marked with a flower in the original score.
In other pieces, the \TT-voice consistently picks the second triad note above the melody or alternates the one above and below it.
\textcite{Hillier1997} called such configurations \emph{tintinnabuli positions}.
As illustrated in \autoref{fig:positions}, he distinguishes two \emph{superior} and two \emph{inferior} positions, which use triad notes \emph{above} and \emph{below} the \MM-voice respectively.
Tintinnabuli positions do not change when transposing them octaves up or down,
and \emph{Für Alina} therefore uses a \TT-voice in first position inferior.

\begin{figure}[t]
    \centering
    \includegraphics[width=\textwidth]{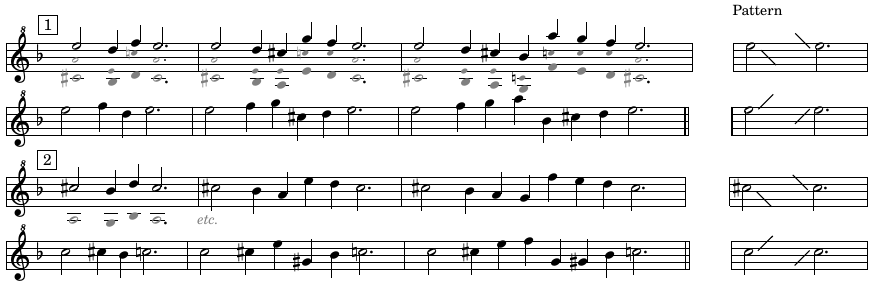}
    \caption{
        \textbf{The melodic structure of \emph{Fratres}} is a series of variations of a six-bar theme, of which the first two are shown.
        The theme leaves and then approaches a pitch center, moving one step further for three consecutive bars.
        The next three bars repeat the first three but played backward.
        This results in four types of melodic movement, or \emph{modes}, that are often used by Pärt to compose \MM-voices.
        The first staff also shows the parallel \MM-voice and the \TT-voice as small notes.
    }
    \label{fig:fratres}
\end{figure}

\begin{figure}[t]
    \centering
    \includegraphics[width=\textwidth]{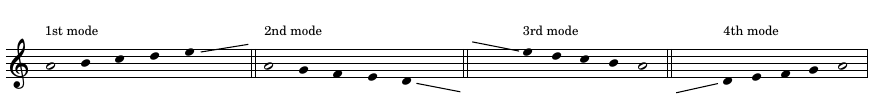}
    \caption{
        \textbf{Four melodic modes} commonly used by Arvo Pärt to construct \MM-voices: two ways to move away from a central pitch, and two to approach it \parencite{Hillier1997}. All of these are found in \emph{Fratres}, as shown in figure \ref{fig:fratres}.
    }
    \label{fig:modes}
\end{figure}

The melody of \emph{Für Alina} seems to be more freely composed than melodies in many of his other works. Still, it follows a numerical regularity: every measure adds another quarter note until the pattern flips midway and measures become shorter and shorter again.
We find an even more systematic melody in
\emph{Fratres} (1977).%
\footnote{
    This analysis is based on the 1980 version for violin and piano.
}
The piece is built around an A minor tintinnabuli triad and has two \MM-voices moving in parallel tenths. 
It is unusually dissonant, as the melodies move along a D harmonic minor scale, which includes a C\musSharp{} instead of the C\musNatural{} from the triad.
\autoref{fig:fratres} illustrates the backbone of \emph{Fratres}: nine variations on a six-measure theme, with each variation lowering the pitch center by another third.
In the first half of the theme, the melody moves down from the pitch center, and then approaches it from above, moving one step further every bar.
The second half of the theme repeats the first half retrogradely.
This results in four types of melodic movement that \textcite{Hillier1997} also frequently encountered in other works of Pärt. 
Figure \ref{fig:modes} summarizes these four melodic \emph{modes}:
moving (1) up or (2) down \emph{from} a central pitch, or moving (3) down or (4) up \emph{towards} it.


The context in which Arvo Pärt developed his tintinnabuli style, and its broader interpretation, has been discussed extensively in the scholarly literature \parencite[see e.g.][]{Hillier1989, Shenton2012, Bouteneff2021}.
It emerged during a period of seven years in which he studied early music, from plainchant to Palestrina, after his earlier serialist style had come to a creative halt.
Musically, as \textcite{Hillier1997} also explains, tintinnabuli contains elements from early polyphony, functional harmony, and serialism.
The stepwise motion in the \MM-voice is for example reminiscent of plainchant, and the homophonic texture it forms with the \TT-voice can be compared to early polyphonic chant settings.
The tintinnabuli style also returns to a form of tonality, but not a functional one.
While the hallmark of function harmony, the triad, is omnipresent in tintinnabuli, it has been stripped of its functional role.
It is remarkable how Pärt managed to fuse all these different ideas into a musical style that appeals to audiences around the world.

\section{Analysis}

We now turn to \emph{Summa}, of which several analyses have been published before.
The first,\footnote{
    \textcite{Shenton2012} also cites the masters thesis by \textcite{Kosak1994}, which I have however not been able to find.
}
\textcite{delaMotte-Haber1996}, preceded Hillier's monograph on Pärt and misses some key points:
it focuses on \TT-voices rather than the \MM-voices and discusses the version for string quartet, in which one cannot see how \emph{Summa} is structured around the text of the Credo, the Christian statement of belief.
\textcite{Hillier1997} points out that syllables in fact form the `units' of the piece and goes on to reveal the structure of the \MM-voices.
But their relation to the \TT-voices remains unclear: although their overall contours correspond, he writes that the ``note-to-note logic of the \TT-voice is, exceptionally, self-contained'' (p.~112).
In an even more extensive analysis, \textcite{Patrick2011} does not resolve this issue either.
The explanation I will propose below indeed moves beyond a note-to-note logic and describes the \TT-voices as tintinnabuli \emph{processes} that also depend on previous notes in the \MM- and \TT-voices. 

But let's start at the beginning.
Hoping to make the text more accessible, I first present an analysis and then a formalization, even though the two developed in tandem and often overlap.

\begin{figure}[t]
    \centering
    \includegraphics[width=\textwidth]{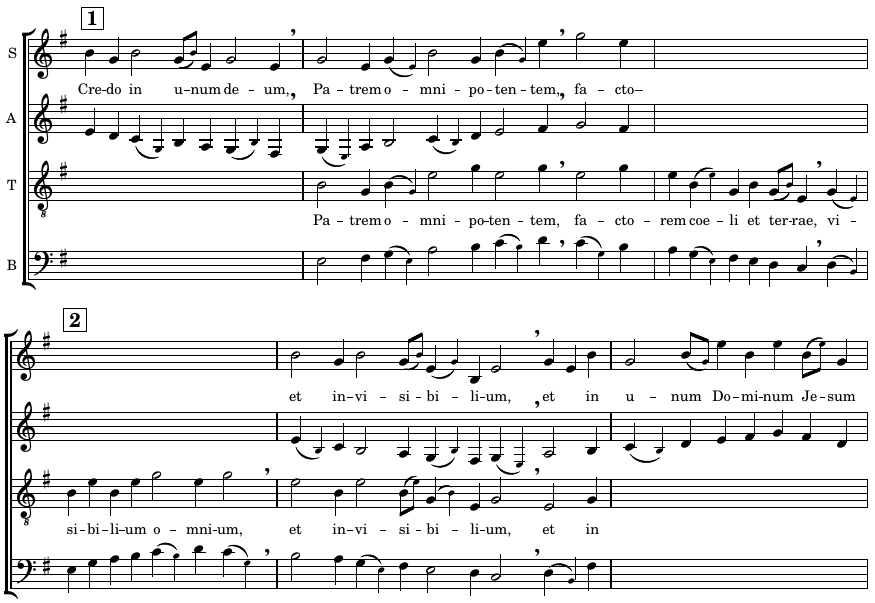}
    \caption{
        \textbf{The opening measures of \emph{Summa}}, a setting of the \emph{Credo}.
        The piece consists of 16 three-bar sections containing 7, 9 and 7 syllables.
        The voice distribution is mirrored in every section: \textsc{sa}--\textsc{satb}--\textsc{tb}.
        The alto and bass have the \MM-voices, the soprano and tenor the corresponding \TT-voices.
        If one skips the small, slurred `ornaments', the \MM-voices walk up and down an E natural minor scale, while the \TT-voices are constrained to the E minor triad. 
    }
    \label{fig:opening}
\end{figure}

\subsection{Text and structure}

Figure \ref{fig:opening} shows the opening bars of \emph{Summa} in the version for mixed choir, which I analyze here.%
\footnote{
    UE 33 686, Korr.~III/2012, to be precise.
}
The first thing that stands out is the overall organization.
\emph{Summa} is divided in 16 sections spanning three measures each.
The first and final bars of a section are sung by the highest (\textsc{sa}) or lowest (\textsc{tb}) two voices, the middle bar is tutti, and this pattern is mirrored in the next section.
The organization becomes transparent when observing that syllables are the unit of time.
The Credo consists of 366 syllables, which Pärt evenly distributed over the 16 sections.
Each section contains 23 syllables, divided over three measures of 7, 9 and 7 syllables respectively (see Appendix \ref{suppl:structure}).
That amounts to a total of 368 syllables, two more than found in the Credo.
The final two bars are composed slightly more freely to 
compensate for this, but as a result break some of the regularities seen in the rest of the score (see Appendix \ref{suppl:final-measures}).

The text setting is homophonic and largely syllabic: most syllables are sung on a single note, some on two notes.
Pärt always slurred those two notes, and we can think of the second one as an \emph{ornament} or passing note \parencite[cf.][]{Hillier1997}.
This distinction between ordinary \emph{notes} and \emph{ornaments} will be important.
The text setting is ``fortuitous'' \parencite{Hillier1997} insofar that it is dictated by the numerical patterns that Pärt laid out, not by the text itself. 
This can be seen in the second bar, where the alto and soprano end without ever finishing the word ``factorem''.
The phrasal structure of the text is maintained in the music and indicated by commas, but these do usually not overlap with bar lines.
A notable exception, as \textcite{Hillier1997} points out, is the very first phrase: ``Credo in unum Deum.''
Its seven syllables may well have inspired the larger structure.

\subsection{Melodic voices}

\begin{figure}
    \includegraphics[width=\textwidth]{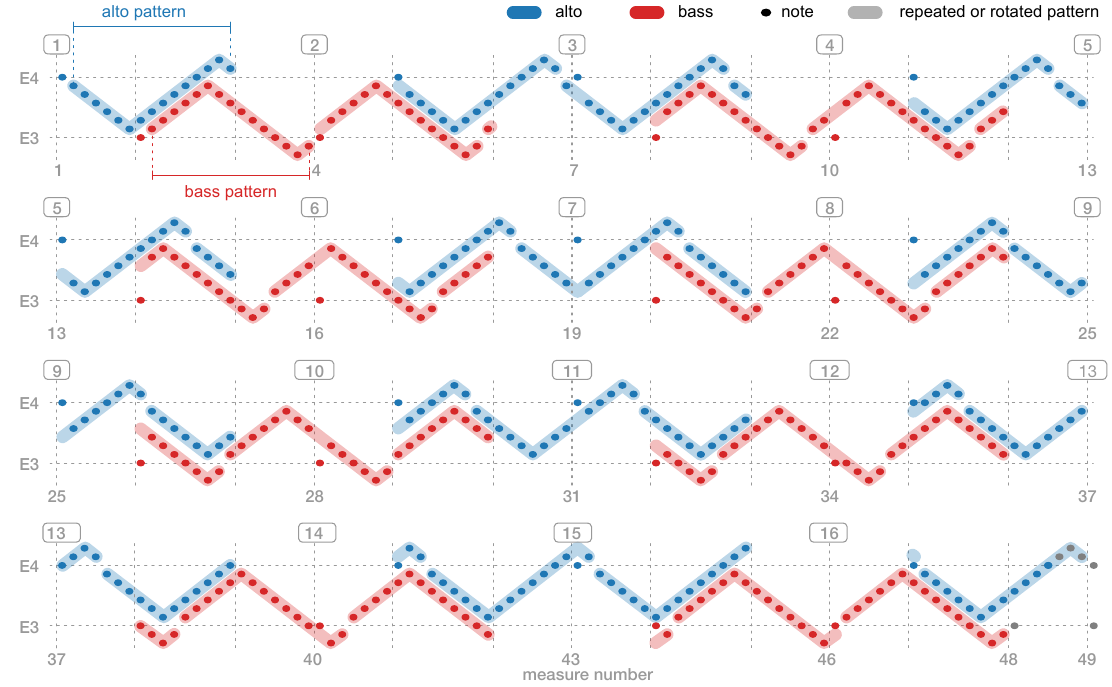}
    \caption{
        \textbf{The melodic voices}
        each repeat a 15-note pattern, that walks up and down a scale, throughout the piece.
        Diatonic pitch is shown vertically and time horizontally, measured in syllables.
        The notes are shown as dots and ornaments have been omitted.
        The repetitions of the underlying 15-note pattern are shown in the background.
        In every section (marked by rehearsal numbers), a voice sings the tonic (E) followed by this pattern, but rotated one step to the left.
    }
    \label{fig:alto-bass}
\end{figure}

\emph{Summa} has two melodic voices: the alto and the bass.
The opening bar makes clear that the soprano is the \TT-voice for the alto, and that the tenor forms a pair with the bass.
Both \TT-voices only sing notes from an E minor triad, and the \MM-voices only use the E natural minor scale, making \emph{Summa} completely diatonic.
The E natural minor scale also forms the backbone of the \MM-voices.
To visualize this, figure \ref{fig:alto-bass} plots only the notes of the alto and bass, and ignores ornaments and note durations.%
\footnote{
    As the piece is diatonic we only have to represent pitches in the E natural minor scale. That means we let $\spn{E}{2}$ correspond to 0, $\text{F}^\#_2$ to 1, $\spn{G}{2}$ to 2, and so on.
}
The blue line highlights that the alto is basically walking up and down the E natural minor scale.
The bass exactly mirrors the alto, but has rests in different places.
Closer inspection shows that the alto is repeating a fifteen-note pattern, which is interrupted by bars of silence and a return to the tonic whenever it enters, or when a new section starts.
As a section contains sixteen syllables, the fifteen-note pattern starts at a different point in every section: it is shifted one step to the left.
And so one can alternatively describe the alto as follows: every section starts with the tonic, followed by the pattern, but rotated one more step to the left \parencite[cf.][]{Patrick2011,Hillier1997}.
Both accounts have the same result, and explain the feeling that the piece could continue forever, were it not for the final two bars.

\subsection{Tintinnabuli voices}

The tintinnabuli voices in \emph{Summa} have puzzled scholars most, even when ignoring ornaments.
Their relation to the \MM-voices is not as direct as in \emph{Für Alina} or \emph{Fratres}, where the \TT-voice consistently takes a fixed tintinnabuli position.
For example, while the first $\spn{C}{4}$ of the alto is paired with a $\spn{B}{4}$ in the soprano, the third time we encounter the $\spn{C}{4}$ in bar 5, the soprano sings a $\spn{G}{4}$.
Both \textcite{Hillier1997} and \textcite{Patrick2011} conclude that the \TT-voices only resemble the shapes of the \MM-voices, but are not predictably related to it.
The \TT-voices indeed cycle through a 30-note pattern that is similarly shaped as the \MM-voices but with slight variations in each repetition.
Still, there appears to be an underlying logic.
To identify it, I `overlaid' all repetitions of this 30-note pattern and worked out an \emph{approximate pattern} that best approximates all of the repetitions (see Appendix \ref{suppl:approximate-patterns}).
Except for a few notes, the approximation will thus be the same as each individual repetition in the score.

\autoref{fig:approximate-pattern} shows the approximate patterns and reveals the constraints that determine the \TT-voices.
First, the soprano is at least two triad notes above the alto and the tenor at least one triad note above the bass.
Second, the \TT-voices only move step-wise to neighboring triad notes (repetitions are not allowed).
It turns out that one obtains the \TT-voices by picking the lowest note satisfying these constraints at every time step.
This also explains some of the slight variations mentioned above: these are caused by the melody voice jumping back to the tonic.
And so the \TT-voices are not in a fixed position but appear to be determined by a \emph{process} that depends on the current melody note and the previous tintinabulli note.

\subsection{Ornaments}

\begin{figure}
    \centering
    \includegraphics[width=\textwidth]{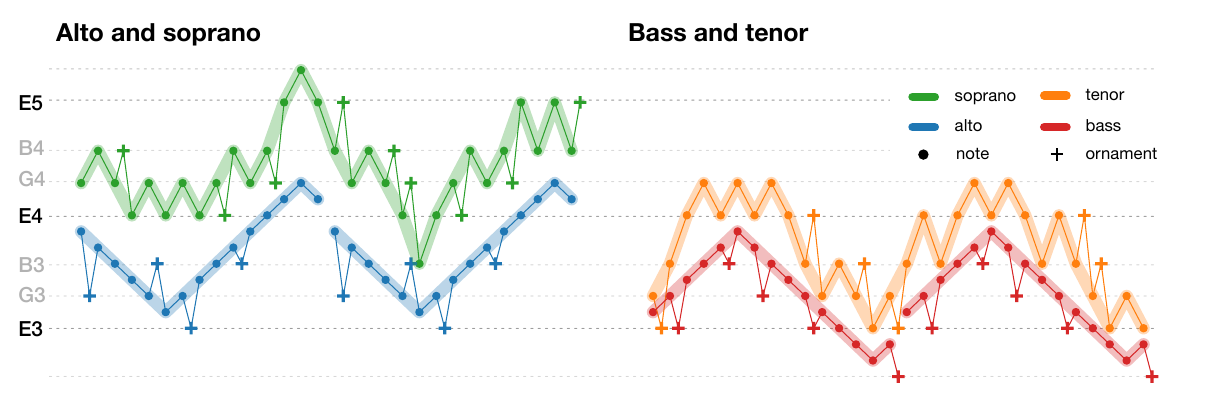}
    \caption{
        \textbf{Approximate patterns} for all voices.
        These patterns of notes (•) and ornaments (+), when repeated throughout the piece, approximate the melodies of each of the voices.
        The approximate patterns were constructed manually by comparing all repetitions, so as to make the approximation as good as possible (see Appendix \ref{suppl:approximate-patterns}).
        For the \TT-voices, they however remain approximations: they are better understood as functions of the \MM-voices (Figure \ref{fig:processes}), not as repetitions of the patterns shown here.
    }
    \label{fig:approximate-pattern}
\end{figure}

P\"art has added ornamental notes to both the \MM- and \TT-voices.
They are always triad notes, which suggests that we can think of the ornaments as tintinnabuli voices themselves.
The approximate patterns in Figure \ref{fig:approximate-pattern} show that ornaments are not randomly inserted, but the underlying pattern is hard to pin down.
For the soprano and tenor we see that ornaments only occur when the melody moves in the same direction for more than two steps, and the ornament reverses the direction.
I see no obvious correspondence between ornaments in the alto and bass pattern.
In particular, they are not mirrored, but they do use the same ornaments (E and B) when passing the G and C on the way up.
That means that for the alto and bass ornaments, the approximate patterns are the best description we currently have.

\subsection{Rhythm}

Finally, the rhythm in \emph{Summa} is determined by two constraints: first, that syllables start together in all voices (homophony), and second, that melody notes have the duration of at least a quarter note.
This implies that if the alto has an ornament where the bass does not, the bass note needs to be twice as long, and vice versa.
If a \TT-voice has two notes where the melody has one quarter, the \TT-voice has to half both of its notes.
These rules are consistently applied throughout the piece, except the penultimate bar, where the bass and tenor start the ``Amen'' before the alto and soprano.

\section{Synthesis}\label{sec:synthesis}

We now formalize the construction of \emph{Summa} laid out above, so that we can implement an algorithm to reconstruct the score.
Our formalism takes inspiration from \textcite{Roeder2011} by distinguishing an \MM- and a \TT-space in which the \MM- and \TT-voices live.
The framework of \textcite{Roeder2011} is so general that it even allows for the possibility that the spaces contain objects other than pitch classes.
That does not help us here: we need to generate specific pitches, and even pitch \emph{classes} would be too general.

Our formalization therefore starts in a larger pitch space $\mathcal N$ that contains all semitones between, say, $\spn{C}0$ and $\spn{C}8$, which are naturally ordered (e.g.~$\spn{G}{2} < \spn{A}{3}$).
If we call a subset $S$ that spans no more than an octave a \emph{scale}, we can generate a \emph{scalar pitch space} $\langle S \rangle$: the pitches (in $\mathcal N$) with the same pitch \emph{class} as elements in the scale.
In this way, we let the E-natural minor scale generate the \MM-space $\mathcal{M}$, in which the \MM-voices can move around.
The \TT-voices live in the \TT-space $\mathcal T$, which is generated by the E minor triad---also a scale under this definition.
In short,
\begin{align}
    \mathcal M &= \langle \noteE, \noteF, \noteG, \noteA, \noteB, \notec, \noted \rangle \\
    \mathcal T &= \langle \noteE, \noteG, \noteB \rangle.
\end{align}
Both spaces are subsets of $\mathcal N$, and $\mathcal T$ is moreover a subset of $\mathcal M$.
But the latter need not be the case: in \emph{Fratres} the triad falls outside of \MM-space because of the C\musNatural{} (\autoref{fig:fratres}).

\subsection{Melodic voices}

We can construct the basic pattern sung by the alto by concatenating fragments of the four melodic modes, or by simply listing its pitches:
\begin{align}
    \alpha 
        &= (\notee, \noted, \notec, \noteB, \noteA, 
        \noteG, \noteF, \noteG, 
        \noteA, \noteB, \notec, \noted, \notee, \notef,
        \noteg, \notef).
\end{align}
The alto sings this 16-note pattern 16 times, but every time rotates the \emph{tail} of the pattern one step to the left: everything after the first note.
For a sequence $x = (x_1, \dots, x_N)$, let $\text{Rotate}(x, d) = (x_{i - d \mod N}: i = 1, \dots, N)$ be its rotation by distance $d$. Then the tail rotation by distance $d$ is $\text{TailRotation}(x, d) = (x_1) \concat \text{Rotate}((x_2, \dots, x_N), d)$, where the cup `$\concat$' indicates concatenation. And so
\begin{align}
    \text{alto} = \text{TailRotation}(\alpha, 0) \; \concat \; \dots \; \concat \;  \text{TailRotation}(\alpha, 15)
\end{align}
gives all notes of the alto.
The bass mirrors this.
Let $\text{mirror}_\mathcal{M}(n, c)$ be the mirror image of $n$ with respect to $c$: the pitch which is equally many steps (in $\mathcal M$) apart from $c$ as $n$ is, but in the other direction.
Then write $\text{transpose}_\mathcal{M}(n, d)$ for the transposition of note $n$ by $d$ steps.
If both these operations work entry-wise on sequences,
\begin{align}
    \text{bass} = \text{transpose}_\mathcal{M}(\text{mirror}_\mathcal{M}(\text{alto}, \spn{E}{4}), -6).
\end{align}

\subsection{Tintinnabuli processes}

The central concept in tintinnabuli music is arguably the tintinnabuli position.
Different from \textcite{Hillier1997}, it will be convenient not to treat positions in octaves as equivalent.
Instead, we denote the tintinnabuli note in $p$-th position above a given note $n$ by $T_p(n)$, and the one below it by $T_{-p}(n)$, and allow $p$ to be any integer.
For example, in our case $T_2(\noteA) = \notee$ since this is the second triad note above $\spn A3$, and $T_{-1}(\noteA) = \noteE$.
One way to define the function $T_p$, is as follows:
\begin{align}
    T_0(n) 
        &= n \\
    T_p(n) 
        &= \min \{ t \in \mathcal{T}: \; t > T_{p-1}(n) \} \\ 
    T_{-p}(n) 
        &= \max \{ t \in \mathcal{T}: \; t < T_{-(p - 1)}(n) \}.
\end{align}
This definition is recursive: we basically think of $T_2(n)$ as the first tintinnabuli note above $T_1(n)$, that is, as $T_1(T_1(n))$.
The function is defined on all of $\mathcal N$, but we are most interested in the case when the position $p$ is nonzero, and $T_p$ maps $\mathcal M$ to $\mathcal T$.

\begin{figure}[t]
    \centering
    \includegraphics[width=\textwidth]{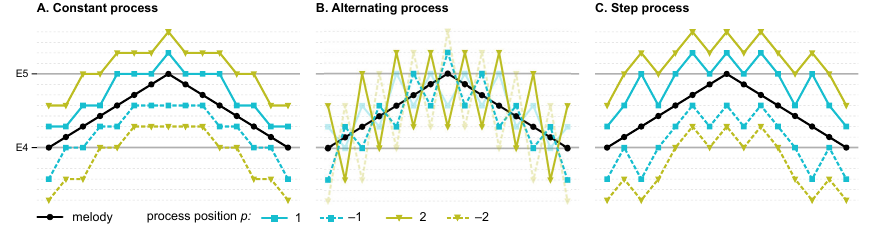}
    \caption{
        \textbf{Three tintinnabuli processes} for a melody that walks up and down a scale.
        The constant process \textbf{(A)} remains in the same tintinnabuli position $p$, while the alternating process \textbf{(B)} flips the sign of the starting position every step.
        The step process \textbf{(C)} only moves to neighbouring triad notes, while keeping a distance of at least $p$ triad notes from the melody. This is the process used in \emph{Summa}.
    }
    \label{fig:processes}
\end{figure}

As we have seen, the tintinnabuli voices in \emph{Summa} are not solely determined by the current melody note, but also by previous notes.
The same is true for Hillier's alternating position (\autoref{fig:positions}).
Because of the sequential dependency, I would propose to speak of a ``process'' instead of a ``position'' in such cases.
To define this formally, consider a sequence of melody pitches $m_1, \dots, m_K$ in $\mathcal M$.
A \emph{tintinnabuli process} $X$ determines a corresponding sequence of tintinnabuli notes $t_1, \dots, t_K$ in $\mathcal T$ via
\begin{align}
    t_{i} = X\bigr((t_1, \dots, t_{i-1}), \; (m_1, \dots, m_K) \bigl), \qquad i = 2, \dots, K.
\end{align}
Such a process can thus depend on all notes in the melody, past and future, but only on previous notes in the tintinnabuli voice.
Of course, we do need to specify the starting point $t_1$, or else the process cannot start.

The simplest example of a tintinabulli process is one that always returns the same tintinabulli position (\autoref{fig:processes}A),
\begin{align}
    \text{Constant}_p(m_i)
        = T_p(m_i).
\end{align}
This shows that a position is a special case of a process. 
A second example would be Hillier's alternating position.
Let $P_m(t)$ denote the position of $t$ with respect to note $m$, that is, the position $p$ such that the $p$-th tintinnabuli note of $m$ is $t$, or $T_p(m) = t$.
Then the alternating process is
\begin{align}
    \text{Alternate}(m_i, m_{i-1}, t_{i-1})
        = T_{-p_i}(m_i),
    \qquad \text{where $p_i = P_{m_{i-1}}(t)$}.
\end{align}
This process basically flips the sign of the starting position (see \autoref{fig:processes}B).

The tintinnabuli voices in \emph{Summa} are determined by a more intricate process that ensures the voices always satisfy two constraints. 
At every point, $t_i$ has to be (1) at least in position $p$ above the melody note $m_i$, and (2) one step in the triad apart from the previous note $t_{i-1}$.
And so the process moves stepwise through T-space while staying at least in position $p$.
This \emph{stepwise tintinnabuli process in position $p$} can be defined as
\begin{align}
    \text{Step}_p(m_i, t_{i-1})
        = \begin{cases}
        T_{-1}( t_{i-1} )  &\text{if this is $\ge T_p(m_{i})$}\\
        T_{+1}( t_{i-1} )  &\text{otherwise}
    \end{cases}, \qquad \text{for $p \ge 0$.}
\end{align}
This process will satisfy both constraints for a stepwise melody as long as the starting point $t_1$ is at least in position $p$.
Although this would also be defined for $p<0$, it seems more appropriate to flip the definition for $p<0$:
\begin{align}
    \text{Step}_p(m_i, t_{i-1})
        = \begin{cases}
        T_{+1}( t_{i-1} )  &\text{if this is $\le T_p(m_{i+1})$}\\
        T_{-1}( t_{i-1} )  &\text{otherwise}
    \end{cases}, \qquad \text{for $p < 0$}.
\end{align}
This resulting process is illustrated in \autoref{fig:processes}C.

\subsection{Ornamentation}

As the ornaments are always triad notes, we can think of the ornaments as \TT-voices, but ones that can also be silent (no ornament), besides singing (an ornament).
This translates into a hierarchy of \TT-voices: the tenor ornaments form a \TT-voice for the tenor, which is a \TT-voice for the bass.
We define a tintinnabuli process that generates the ornaments for both the soprano and the tenor (see Appendix \ref{suppl:repeat-previous}).
The process returns the previous melody note if it does not equal the next melody note (the \MM- and \TT-spaces are identical), and while it remains within certain bounds:
\begin{align}
    \text{RepeatPrevious}_{b,B,c,C}(m_{i-1}, m_{i+1}) =
        \begin{cases}
                       & \text{when $m_{i+1} \neq m_{i-1}$,}\\    
        m_{i-1}        & \text{and $b \le m_{i-1} \le B$}\\
                       & \text{and $c \le m_{i+1} \le C$}\\    
        \text{silent}  & \text{otherwise}.
        \end{cases}
\end{align}
Without the bounds, the process cannot avoid ornaments at the extremes of the range (e.g., $\spn{G}{5}$ or $\spn{B}{3}$ for the soprano), and one can imagine why Pärt might have wanted to avoid those.
Although it remains a question whether Pärt actually thought of the ornamentation in this way, the reuse of formal machinery seems appealing.

Finally, the alto and bass have ornaments at fairly regular positions along the 16-note melodic pattern.
We therefore define a process that repeats a fixed sequence of ornamental pitches $x$, which can also contain silences.
Since the melodic pattern is repeated with a tail rotation, we also need to rotate $x$ to keep it aligned:
\begin{align}
    \text{TailRotatedPattern}_x(m_i) = r_{i \mod |x|}, 
    \; \text{where} \; r = \text{TailRotation}(x, \text{floor}(i / |x|)).
\end{align}
The pattern of ornaments we use is illustrated in Figure \ref{fig:approximate-pattern} (and in Appendix \ref{suppl:tail-rotated-pattern}).

\subsection{Implementation}

To summarize, our formalism describes the notes of the alto as a tail-rotated pattern and the bass as its mirror image.
The soprano and tenor are stepwise tintinnabuli processes in second and first position respectively. 
Ornaments are also described as tintinnabuli processes.
We can then insert the notes and ornaments into the measure structure discussed in the previous section, and determine the note duration.
For the latter, we first assign every syllable a duration: $2$ if either the alto or the bass has an ornament, and $1$ otherwise.
Then we evenly distribute the available time over the notes of a voice.
I implemented all this in Python using the computational musicology package music21 \parencite{Cuthbert2010}.
The codebase, named \emph{tintinnabulipy},%
\footnote{
    All code used in this paper can be found on \url{https://github.com/bacor/algo-part}
}
provides a convenient interface for plotting and working with \TT- and \MM-spaces.
It implements all of the tintinnabuli processes described here but is also general enough to be useful for analyses of other compositions by Pärt.
Most importantly, it allowed me to generate almost all melodic material of \emph{Summa} in just a few lines of code (see Appendix \ref{suppl:code-sample}).

\section{Evaluation}

\begin{figure}
    \footnotesize
    \textbf{A. Types of errors in the reconstruction}
    \hspace{4cm}
    \textbf{B. Error frequencies}\\[1em]
    \includegraphics[width=9cm]{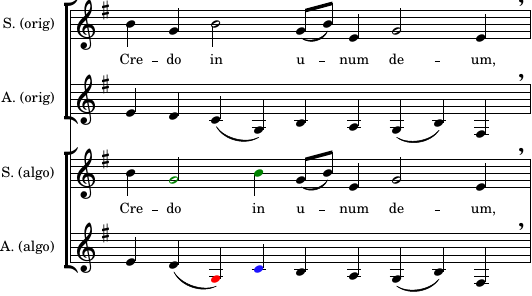}
    \hfill
    \includegraphics[width=5cm]{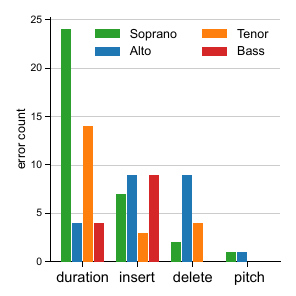}
    \caption{
        \textbf{Evaluation of the algorithmic reconstruction} 
        illustrated in \textbf{(A)} by comparing the first bar (bottom staves) with the original (top staves). 
        We encounter four types of reconstruction errors: ornament insertions (red), ornament deletions (blue), duration errors (green), and pitch errors (not shown).
        In total, we find 86 errors (6.7\%) after adjusting ornamented exits \textbf{(B)}.
        Over half of these are duration errors, resulting from ornament insertions or deletions.
        And so only 43 ornaments and two pitches need to be corrected (3.5\%) to reproduce the original score.
    }
    \label{fig:evaluation}
\end{figure}

How much of the original composition is reproduced by our algorithm? 
\autoref{fig:evaluation}A compares the first bar of the original score with the algorithmic reconstruction. 
The reconstruction contains four mismatches---I will call these \emph{errors} for simplicity---in the second and third syllable, which have been colored according to their type.
First, we see an \emph{ornament insertion} in the third note of the reconstructed alto part: the reconstruction has an ornament, but the original does not.
Conversely, an \emph{ornament deletion} occurs when the original is ornamented, but the reconstruction is not, as with the fourth note in the alto.
We also see several \emph{duration errors}:
the second note of the soprano for example has double the duration of the original. 
Finally, \emph{pitch errors} occur when a note has the wrong pitch, but these do not appear in this excerpt.
I automated this evaluation to systematically compare the reconstructed score with the original score, part by part and syllable by syllable.%
\footnote{
    To obtain a digital version of the original score, I transcribed my physical copy in MuseScore.
    I manually compared all errors identified in reconstruction against the physical score, and this allowed me to resolve some transcription mistakes.
}

The reconstructed score contains 1288 notes, of which 106 (8\%) have one or more errors.
Most errors only concern the note duration (60 notes or 56\%), but we also find 2 pitch errors, 15 ornament deletions, and 34 insertions, eight of which are in the final two bars.
These results show that our algorithmic reconstruction is fairly successful: it correctly reproduces well over 90\% of the notes in \emph{Summa}.
And this statistic arguably underestimates the performance, since all duration errors are explained by ornament insertions or deletions.
If the alto for example misses an ornament, this causes the corresponding soprano note to be too short.
And so fixing insertion and deletion errors will automatically resolve all duration errors.
That means that only 51 notes (4\%) in the reconstruction really need to be corrected in order to reproduce the original score faithfully.

The remaining errors however reveal another plausible regularity.
In the reconstruction, one finds several  ornaments right before a voice exits to be silent for some measures, whereas ornamented exits are not found in the original score.
Removing all ornamented exits resolves six insertions and consequently also reduces the number of duration errors, leaving a total of 86 errors (7\%).
Of these, 45 (3.5\%) are not duration errors and need to be corrected.
The alto needs the most correction (19 notes) and is around twice as inaccurate as the soprano, tenor, and bass (10, 7, and 9).
This is also summarised in \autoref{fig:evaluation}B.
Taking into account that eight errors occur in the final bars, and many other errors remain in the ornamentation, the reconstruction seems very accurate and underscores just how meticulously Pärt constructed \emph{Summa}.

\section{Discussion and conclusion}

Arvo Pärt is known for his unique compositional style, \emph{tintinnabuli}, which has often been described as algorithmic.
To assess \emph{how} algorithmic Pärt's tintinnabular music is, this study has attempted to reconstruct one piece, \emph{Summa}, algorithmically.
After analyzing and formalizing the piece, I arrived at an implementation that reconstructed most of the original score, showing that at least 93\% of the notes in \emph{Summa} can be plausibly explained by an algorithm.
Most of the errors, moreover, are faulty note durations caused by insertions or deletions of ornaments in other voices.
Correcting these ornamental errors would also resolve the duration errors. 
This means that only 3,5\% of the notes have to be corrected to retrieve the original score, and demonstrates that Arvo Pärt approached the composition of \emph{Summa} extremely systematically.

One might wonder whether the algorithm that I proposed also describes the compositional process: were these the procedures Pärt followed?
That may seem plausible, but only the composer can answer that question and Pärt is unlikely to do so.%
\footnote{
    In the comments that are reproduced in footnote \ref{fn:parts-comment}, Pärt expresses his ``great difficulty'' in commenting on his own works as he wants to give the music a chance to express itself.
}
If my analysis is mistaken, the mistakes are probably in the description of the ornamentation, where we found the most errors.
However, we should also consider that the composer may have decided to adjust some of the ornaments and that there are no further regularities to be found.
After all, multiple corrections of the score have been published.
Although I have not been able to compare all editions, some differences in ornamentation can also be heard in recordings.\footnote{
    To give just one example, in measure 12 the score used in this study ornaments 've-' in 'verum' with an $\spn{E}{4}$ in the alto, while the recording by the Hilliard Ensemble (1987) sings a $\spn{G}{4}$, as does the more recent recording by Vox Clamantis (2016).
}
These corrections also leave open the curious possibility that the composer has made `mistakes' when applying his set of rules. 
Doing so by hand, rather than by computer, is far from straightforward and would be comparable to a composer from earlier days making an occasional mistake in voice leading.

While analyzing \emph{Summa}, I developed some novel formal machinery.
Most notably, I proposed \emph{tintinnabuli processes} to describe how a \TT-voice can be produced from an \MM-voice while relying on parts of the melody other than the current melody note. 
This turned out to be a fruitful generalization of Hillier's tintinnabuli positions.
I expect other analyses will also benefit from this concept---as they will from formalization more generally: the intricacies of works like \emph{Summa} are arguably best described in a formal language.
This study demonstrates that it can be useful to also implement that formalism, and I hope the resulting codebase will contribute to further formal and computational analyses of Pärt's work.

The methodology this study proposed for that, \emph{analysis by synthesis}, is best suited for understanding algorithmic music: it essentially tries to recover the rules that generated a piece.
But it could have wider applicability.
Strictly speaking, any piece can be algorithmically reconstructed by simply enumerating all notes in the score.
The more rules a piece satisfies, the more concise the description can be.
Algorithmic music is an extreme case, but other types of music also follow rules.
It may well be possible to for example recover fragments of the middle voices in a Bach chorale from the melody, a figured bass, and voice-leading rules.

That is not to say that algorithmic reconstruction should replace other forms of scholarship.
This study has deliberately disregarded all matters of interpretation, which are of course central to understanding the music of Pärt in a broader sense.
For that, a methodology like analysis by synthesis seems less useful.
But when it comes to understanding how Arvo Pärt's tintinnabular compositions work, this study may provide a fruitful starting point.

\section*{Acknowledgements}

I would like to thank Dorien Salet for introducing me to \emph{Summa} and the members of ensemble Bes Klein for giving me a reason to analyse the piece in detail.
This would not have happened if Euwe de Jong had not first introduced me to the tintinnabuli style.
Finally, I would like to thank Ashley Burgoyne and Jelle Zuidema for their comments on the manuscript, and Iris Bouman for helping me with some German translations.

\theendnotes

\printbibliography
\newpage\appendix

\section{Textual structure}
\label{suppl:structure}

\begin{table}[h!]
    \footnotesize
    \setlength{\tabcolsep}{2pt}
    \begin{tabular}{lllllllllllll}
    \toprule
    Section & Bar &Voices & Syllables & Text\\
    \midrule
    \framebox{\tiny1}
	&1  &\textsc{sa}   &7 &Cre-  &do    &in    &u-    &num   &de-   &um,   \\
   	&2  &\textsc{satb} &9 &Pa-   &trem  &o-    &mni-  &po-   &ten-  &tem,  &fa-   &cto-  \\
   	&3  &\textsc{tb}   &7 &rem   &coe-  &li    &et    &ter-  &rae,  &vi-   \\
\framebox{\tiny2}
	&4  &\textsc{tb}   &7 &si-   &bi-   &li-   &um    &o-    &mni-  &um,   \\
   	&5  &\textsc{satb} &9 &et    &in-   &vi-   &si-   &bi-   &li-   &um,   &et    &in    \\
   	&6  &\textsc{sa}   &7 &u-    &num   &Do-   &mi-   &num   &Je-   &sum   \\
\framebox{\tiny3}
	&7  &\textsc{sa}   &7 &Chri- &stum, &Fi-   &li-   &um    &De-   &i     \\
   	&8  &\textsc{satb} &9 &u-    &ni-   &ge-   &ni-   &tum,  &et    &ex    &Pa-   &tre   \\
   	&9  &\textsc{tb}   &7 &na-   &tum   &an-   &te    &o-    &mni-  &a     \\
\framebox{\tiny4}
	&10 &\textsc{tb}   &7 &sae-  &cu-   &la.   &De-   &um    &de    &De-   \\
   	&11 &\textsc{satb} &9 &o,    &lu-   &men   &de    &lu-   &mi-   &ne,   &De-   &um    \\
   	&12 &\textsc{sa}   &7 &ve-   &rum   &de    &De-   &o     &ve-   &ro,   \\
\framebox{\tiny5}
	&13 &\textsc{sa}   &7 &ge-   &ni-   &tum,  &non   &fa-   &ctum, &con-  \\
   	&14 &\textsc{satb} &9 &sub-  &stan- &ti-   &a-    &lem   &Pa-   &tri:  &per   &quem  \\
   	&15 &\textsc{tb}   &7 &o-    &mni-  &a     &fac-  &ta    &sunt. &Qui   \\
\framebox{\tiny6}
	&16 &\textsc{tb}   &7 &prop- &ter   &nos   &ho-   &mi-   &nes,  &et    \\
   	&17 &\textsc{satb} &9 &pro-  &pter  &no-   &stram &sa-   &lu-   &tem   &de-   &scen- \\
   	&18 &\textsc{sa}   &7 &dit   &de    &coe-  &lis.  &Et    &in-   &car-  \\
\framebox{\tiny7}
	&19 &\textsc{sa}   &7 &na-   &tus   &est   &de    &Spi-  &ri-   &tu    \\
   	&20 &\textsc{satb} &9 &San-  &cto   &ex    &Ma-   &ri-   &a     &Vir-  &gi-   &ne:   \\
   	&21 &\textsc{tb}   &7 &Et    &ho-   &mo    &fa-   &ctus  &est.  &Cru-  \\
\framebox{\tiny8}
	&22 &\textsc{tb}   &7 &ci-   &fi-   &xus   &e-    &ti-   &am    &pro   \\
   	&23 &\textsc{satb} &9 &no-   &bis   &sub   &Pon-  &ti-   &o     &Pi-   &la-   &to    \\
   	&24 &\textsc{sa}   &7 &pas-  &sus   &et    &se-   &pul-  &tus   &est.  \\
\framebox{\tiny9}
	&25 &\textsc{sa}   &7 &Et    &re-   &sur-  &re-   &xit   &ter-  &ti-   \\
   	&26 &\textsc{satb} &9 &a     &di-   &e,    &se-   &cun-  &dum   &scri- &ptu-  &ras.  \\
   	&27 &\textsc{tb}   &7 &Et    &a-    &scen- &dit   &in    &coe-  &lum,  \\
\framebox{\tiny10}
	&28 &\textsc{tb}   &7 &se-   &det   &ad    &dex-  &te-   &ram   &Pa-   \\
   	&29 &\textsc{satb} &9 &tris. &Et    &i-    &te-   &rum   &ven-  &tu-   &rus   &est   \\
   	&30 &\textsc{sa}   &7 &cum   &glo-  &ri-   &a,    &ju-   &di-   &ca-   \\
\framebox{\tiny11}
	&31 &\textsc{sa}   &7 &re    &vi-   &vos   &et    &mor-  &tu-   &os,   \\
   	&32 &\textsc{satb} &9 &cu-   &jus   &re-   &gni   &non   &e-    &rit   &fi-   &nis.  \\
   	&33 &\textsc{tb}   &7 &Et    &in    &Spi-  &ri-   &tum   &San-  &ctum, \\
\framebox{\tiny12}
	&34 &\textsc{tb}   &7 &Do-   &mi-   &num,  &et    &vi-   &vi-   &fi-   \\
   	&35 &\textsc{satb} &9 &can-  &tem:  &qui   &ex    &Pa-   &tre   &Fi-   &li-   &o-    \\
   	&36 &\textsc{sa}   &7 &que   &pro-  &ce-   &dit.  &Qui   &cum   &Pa-   \\
\framebox{\tiny13}
	&37 &\textsc{sa}   &7 &tre   &et    &Fi-   &li-   &o     &si-   &mul   \\
   	&38 &\textsc{satb} &9 &ad-   &o-    &ra-   &tur,  &et    &con-  &glo-  &ri-   &fi-   \\
   	&39 &\textsc{tb}   &7 &ca-   &tur,  &qui   &lo-   &cu-   &tus   &est   \\
\framebox{\tiny14}
	&40 &\textsc{tb}   &7 &per   &Pro-  &phe-  &tas.  &Et    &u-    &nam   \\
   	&41 &\textsc{satb} &9 &san-  &ctam  &ca-   &tho-  &li-   &cam   &et    &a-    &po-   \\
   	&42 &\textsc{sa}   &7 &sto-  &li-   &cam   &Ec-   &cle-  &si-   &am.   \\
\framebox{\tiny15}
	&43 &\textsc{sa}   &7 &Con-  &fi-   &te-   &or    &u-    &num   &ba-   \\
   	&44 &\textsc{satb} &9 &pti-  &sma   &in    &re-   &mis-  &si-   &o-    &nem   &pec-  \\
   	&45 &\textsc{tb}   &7 &ca-   &to-   &rum.  &Et    &ex-   &spe-  &cto   \\
\framebox{\tiny16}
	&46 &\textsc{tb}   &7 &re-   &sur-  &re-   &cti-  &o-    &nem   &mor-  \\
   	&47 &\textsc{satb} &9 &tu-   &o-    &rum,  &et    &vi-   &tam   &ven-  &tu-   &ri    \\
   	&48 &\textsc{sa}   &4 &sae-  &cu-   &li.   &A-    \\
	&49 &\textsc{satb} &1 &men   \\
    \bottomrule
    \end{tabular}
    \caption{
        \textbf{Structure of \emph{Summa}.}
        The piece consists of 16 sections, marked by rehearsal numbers, of each 3 bars.
        Measures in a section contain 7, 9 and 7 syllables respectively, and use different voices:  \textsc{sa}, then \textsc{satb} and finally \textsc{tb}. 
        The next section mirrors this structure.
        Syllables are distributed across the bars following this scheme, even if this means that a bar line falls in the middle of a word.
    }
    \label{tab:architecture}
\end{table}

\pagebreak\newpage

\section{Approximate patterns}\label{suppl:approximate-patterns}

\begin{figure}[h!]
    \includegraphics[width=\textwidth]{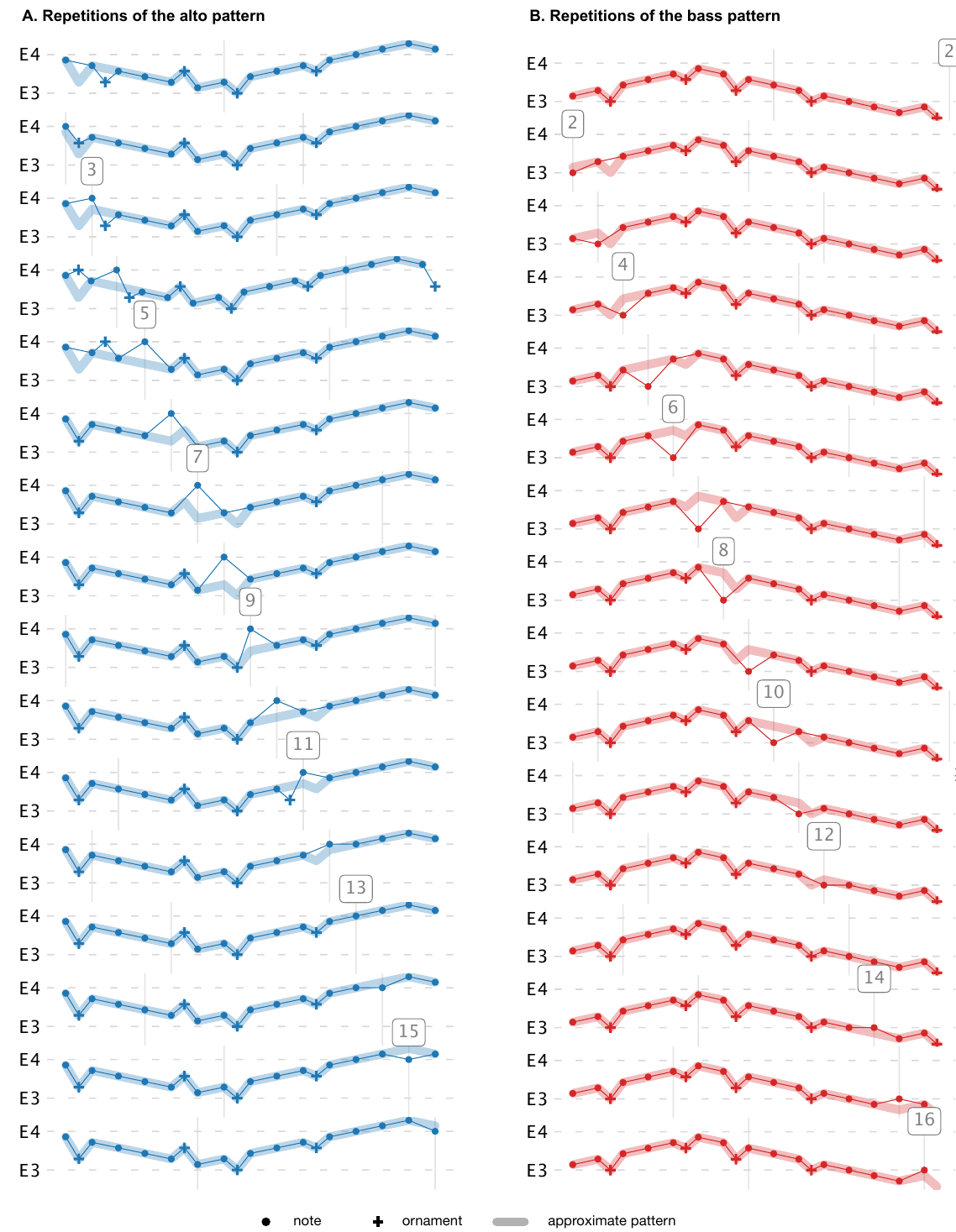}
    \caption{
        \textbf{Pattern repetitions of the melodic voices.}
        All repetitions of the alto (A) and bass (B) are plotted above one another. We manually identified an \emph{approximate pattern} of notes and ornaments (shown in the background) that best matches all of the repetitions. In other words, it minimizes the number of deviations. For the \TT-voices (Figure \ref{fig:repetitions-t-voices}), this turned out to be a crucial step in understanding their construction.
    }
    \label{fig:repetitions-m-voices}
\end{figure}

\newpage

\begin{figure}[h!]
    \includegraphics[width=\textwidth]{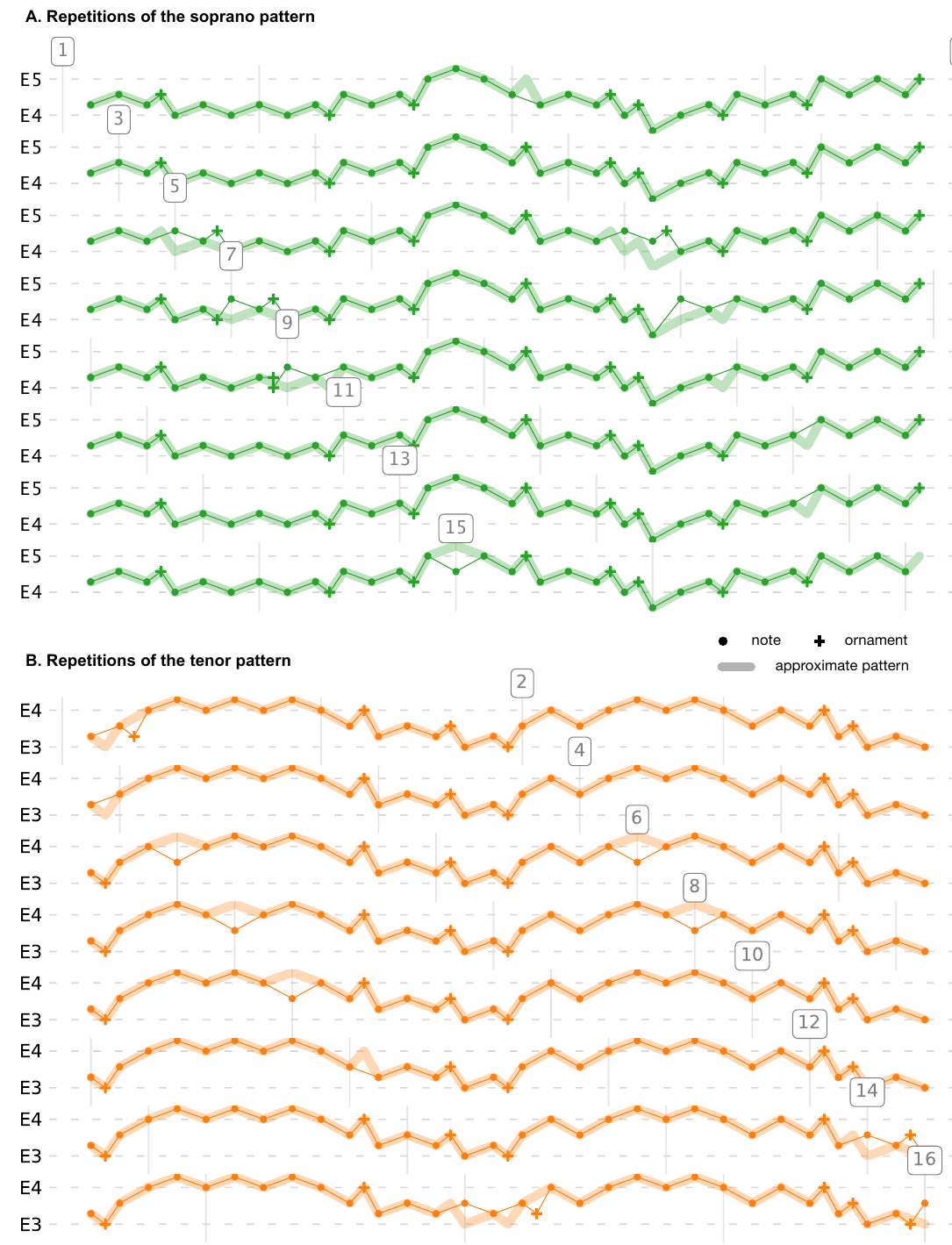}
    \caption{
        \textbf{Pattern repetitions of the tintinnabuli voices.}
        This is very similar to Figure \ref{fig:repetitions-m-voices}, but now the patterns are twice as long.
    }
    \label{fig:repetitions-t-voices}
    \vfill
\end{figure}

\pagebreak\newpage

\section{Tenor and soprano ornament processes}\label{suppl:repeat-previous}

\begin{figure}[h!]
    \includegraphics[width=16cm]{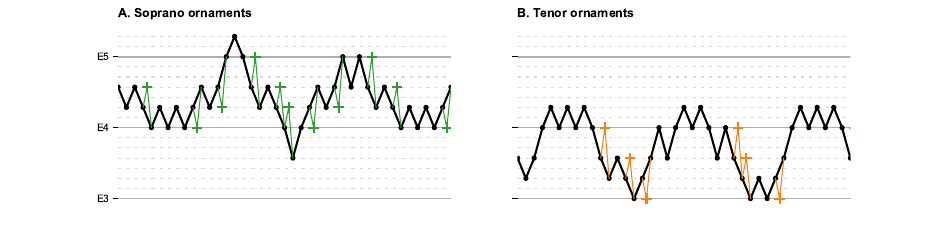}
    \caption{
        \textbf{The RepeatPrevious process}
        generates the ornaments for the soprano \textbf{(A)} and tenor \textbf{(B)} by repeating the previous note if this is not equal to the next note.
        As explained in the main text, this process has several parameters that constrain the range of the ornaments.
        For the soprano the ornaments have to lie between $b=\spn{E}{4}$ and $B=\spn{E}{5}$, while the next note has to  fall below $C=\spn{E}{5}$.
        For the tenor, we use $b=\spn{E}{3}$, $B=\spn{E}{4}$ and $C=\spn{B}{3}$.
    }
    \label{fig:repeat-previous}
\end{figure}

\section{Alto and bass ornament processes}\label{suppl:tail-rotated-pattern}

\begin{figure}[h!]
    \centering
    \includegraphics[width=16cm]{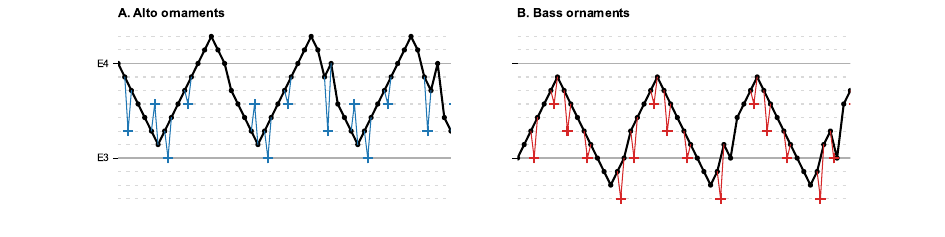}
    \caption{
        \textbf{The TailRotatedPatternProcess process}
        generates the ornaments for the alto \textbf{(A)} and bass \textbf{(B)} parts.
        The black lines show the respective melodies, and ornaments are indicated by coloured plusses.
        It essentially repeats a 16-note pattern of ornamentation, but rotates the tail every time to keep the ornamentation in sync with the melody (see main text for details).
    }
    \label{fig:my_label}
\end{figure}

\pagebreak\newpage

\section{Implementation: code sample}\label{suppl:code-sample}

\begin{figure}[h!]
    \centering\footnotesize
    \begin{verbatim}
# Define the melodic spaces
M = MelodicSpace(MinorScale('E4'))
T = TintinnabuliSpace(Chord(['E4', 'G4', 'B4']))

# Construct the alto melody and ornaments
alto_pattern = glue(M.mode2(6), M.mode4(6), M.mode1(2), M.mode3(2))[:-1]
repetitions = [rotate_tail(alto_pattern, i) for i in range(16)]
alto = concatenate(*repetitions)
ornament_pattern = [None, 'G3', None, None, None, 'B3', None, 'E3', 
                    None, None, 'B3', None, None, None, None, None]
alto_orn_process = TailRotatedPatternProcess(T, alto_pattern)
alto_ornaments = alto_process(alto, t0=False)

# Construct the soprano melody and ornaments
soprano = StepProcess(T, position=2)(alto)
sop_orn_process = RepeatPreviousProcess(T, ['E4', 'E5'], [None, 'E5'])
sop_ornaments = sop_orn_process(soprano, t0=soprano[0])
\end{verbatim}
    \caption{\textbf{Fragment of the implementation.} Using tintinnabulipy, all notes and ornaments of the alto and soprano can be constructed in just a few lines of code. The tenor and bass are similar. The majority of the remaining code is needed to turn this into an actual score (i.e., a musicxml file).
    }
    \label{fig:code-fragment}
\end{figure}

\pagebreak\newpage

\section{Ending of \emph{Summa}}\label{suppl:final-measures}
\begin{figure}[h!]
    \centering
    \includegraphics[width=\textwidth]{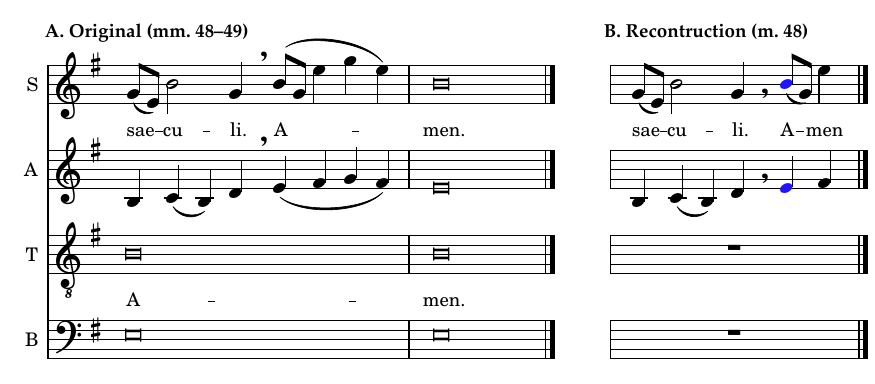}
    \caption{
        \textbf{Final bars of \emph{Summa}.}
        The ending of \emph{Summa} \textbf{(A)} is more freely composed than the rest of the piece and deviates from the patterns observed before.
        The alto finishes the last repetition of the basic pattern in a four-note melisma, to end on the tonic.
        Meanwhile, the bass and tenor hold an open fifth on 'Amen'.
        The reconstruction \textbf{(B)} of course cannot accurately reproduce these measures.
        We treat the errors as ornament insertions, and count 2 extra insertions for all voices.
    }
    \label{fig:summa-ending}
\end{figure}

\end{document}